\documentstyle[latexsym,amssymb,epsfig,aps,floats,eqsecnum,amsfonts,preprint]{revtex}

\def\laq{\raise 0.4ex\hbox{$<$}\kern -0.8em\lower 0.62ex\hbox{$\sim$}}
\def\gaq{\raise 0.4ex\hbox{$>$}\kern -0.7em\lower 0.62ex\hbox{$\sim$}}

\def\fun#1#2{\lower3.6pt\vbox{\baselineskip0pt\lineskip.9pt
\ialign{$\mathsurround=0pt#1\hfil##\hfil$\crcr#2\crcr\sim\crcr}}}

\newcommand{\beq}{\begin{equation}} 
\newcommand{\eeq}{\end{equation}}
\newcommand{\bea}{\begin{eqnarray}} 
\newcommand{\eea}{\end{eqnarray}}
\newcommand{\vphi}{\varphi}

\newcommand{\be}{\mbox{\boldmath${e}$}}
\newcommand{\ba}{\mbox{\boldmath${a}$}}
\newcommand{\bv}{\mbox{\boldmath${v}$}}
\newcommand{\bP}{\mbox{\boldmath${P}$}}
\newcommand{\br}{\mbox{\boldmath${r}$}}
\newcommand{\bz}{\mbox{\boldmath${z}$}}
\newcommand{\bq}{\mbox{\boldmath${q}$}}
\newcommand{\bqd}{\dot{\mbox{\boldmath${q}$}}}
\newcommand{\bR}{\mbox{\boldmath${R}$}}

\newcommand{\bJ}{\mbox{\boldmath${J}$}}
\newcommand{\bcJ}{\mbox{\boldmath${\cal J}$}}
\newcommand{\cJ}{{\cal J}}
\newcommand{\bj}{\mbox{\boldmath${j}$}}
\newcommand{\bA}{\mbox{\boldmath${A}$}}
\newcommand{\bp}{\mbox{\boldmath${p}$}}
\newcommand{\bn}{\mbox{\boldmath${n}$}}
\newcommand{\cL}{{\cal L}} 
\newcommand{\cI}{{\cal I}} 
\newcommand{\cN}{{\cal N}} 
\newcommand{\cH}{{\cal H}} 
\newcommand{\cE}{{\cal E}}

\begin{document}
\preprint{\vbox{\baselineskip=12pt
\rightline{GRP/00/523} 
\vskip 0.2truecm
\rightline{hep-th/0004042}}}

\title{\Large\bf Reduction of the two-body dynamics 
to a one-body description in classical electrodynamics}
\author{Alessandra Buonanno}

\address{\it Theoretical Astrophysics and Relativity Group \\ 
California Institute of Technology, Pasadena, CA 91125, USA }

\maketitle
\begin{abstract}
We discuss the mapping of the conservative part of two-body  
electrodynamics onto that of a test charged particle 
moving in some external electromagnetic field,  
taking into account recoil effects and relativistic 
corrections up to second 
post-Coulombian order. Unlike the results recently 
obtained in general relativity, we find that 
in classical electrodynamics 
it is not possible to implement 
the matching without introducing external 
parameters in the effective electromagnetic field.  
Relaxing the assumption that the effective test 
particle moves in a flat spacetime provides a 
feasible way out. 
\end{abstract}
\newpage
\section{Introduction}
\label{sec1}

Recently, a novel approach to the two-body problem 
in general relativity has been introduced~\cite{BD99}. 
The main motivation of that investigation 
rests on better understanding  
the late dynamical evolution of a coalescing 
binary system made of compact bodies of comparable masses, 
such as black holes and/or neutron stars. In fact, 
these astrophysical systems are among 
the most promising candidate sources for the detection 
of gravitational-waves with the future terrestrial interferometers 
such as the Laser Interferometric Gravitational 
Wave Observatory (LIGO) and Virgo. The basic idea pursued in \cite{BD99}, 
in part inspired by some results obtained in quantum 
electrodynamics~\cite{BIZ70,T70}, was 
to map the conservative two-body dynamics (henceforth 
denoted as the ``real'' dynamics)  onto an effective 
one-body one, where a test particle moves in an effective external metric. 
As long as radiation reaction effects are 
not taken into account, the effective metric 
is just a deformation of the Schwarzschild 
metric with deformation parameter $\nu = \mu/M$, 
where $\mu$ is the reduced mass of the binary system and 
$M$ its total mass. The ``effective'' description  
should be viewed as a way of re-summing in a non-perturbative 
manner the badly convergent post-Newtonian-expanded dynamics 
of the ``real'' description. The results in 
\cite{BD99} were restricted to the 
second post-Newtonian level (2PN) and the analysis was 
mainly focused on the conservative part of the dynamics.
More recently, a feasible way of incorporating radiation 
reaction effects has been proposed \cite{BD00} and 
the extension of the aforesaid approach to 
3PN order has been investigated~\cite{DJS00}.

The purpose of the present paper is to test the robustness 
of the basic idea underlying the mapping of the two-body 
problem onto an effective one-body one, by applying it 
to classical electrodynamics. We limit 
to the conservative part of the 
dynamics of the bound states of 
two charged particles, up to second post-Coulombian order (2PC),
and we take into account recoil effects. We investigate 
the possibility of describing the exchange of 
energies between the two bodies in the ``real'' 
problem through an ``effective'' auxiliary 
description, where a test particle moves 
in some external effective electromagnetic field. 
Generically, we expect that this electromagnetic 
field will be a deformation of the Coulomb potential with deformation 
parameter $\nu = \mu/M$, where $\mu$ is 
the usual reduced mass of the two charged 
particles and $M$ the total mass of the system.
We shall see that the matching  
is also possible introducing in the 
effective description either a $\nu$-dependent vector 
potential or a deformed flat metric with 
deformation parameter $\nu$.  

As already mentioned, the idea of reducing the relativistic two-body dynamics 
onto a relativistic one-body one was originally introduced in 
quantum electrodynamics. 
In particular, in \cite{BIZ70} the authors, taking into 
account recoil effects, resummed in the eikonal approximation 
the ``crossed-ladder'' Feynman diagrams for the scattering of two
relativistic particles and mapped 
the one-body relativistic Balmer formula onto   
the two-body relativistic one. This method gives the correct quantum energy 
levels at least up to 1PC order, but some of the centrifugal barrier effects 
have to be added by hand. Todorov et al.~\cite{T70} developed a more 
systematic approach, based on the Lyppmann-Schwinger 
quasi-potential equation, which also gives correct results 
for the quantum energy levels, including the 
main parts of the radiative effects of the Lamb shift. Nevertheless, this last  
approach \cite{T70} rests on some choices for the quasi-potential 
equation which are not very well justified and introduces in the effective description   
various energy-dependent quantities. In the following, whenever it is
possible, we will compare our results in classical electrodynamics 
with the previous analysis for the corresponding quantum problem.
Finally, note that, the aim of this paper is not to obtain 
new results with respect to the quantum energy-levels of the bound states 
of a two-body charged system, which is well known to be a hard problem 
\cite{Lit}. On the other hand, the present work wants to  
investigate, in the context of classical electrodynamics, 
the basic idea of reducing  the two-body 
dynamics onto a one-body one, recently introduced 
in general relativity \cite{BD99}. 

The outline of the paper is as follows. In Section~\ref{sec2} 
we review the relativistic two-body problem up to 
2PC order and summarize its dynamics in a coordinate-invariant manner 
evaluating , within the Hamilton-Jacobi framework, the 
``energy-levels'' of the bound states.
In Section~\ref{sec3} we introduce the ``effective'' one-body 
description and define the ``rules'' needed to map 
the ``real'' onto the ``effective'' problem. Then, 
in Sections \ref{subsec3.1}, \ref{subsec3.2} and \ref{subsec3.3} 
we analyze three feasible manners of implementing the matching.  
Finally, Section \ref{sec4} summarizes our main 
conclusions.

\section{Two-body dynamics up to second post-Coulombian order}
\label{sec2}

It was realized long ago that, in relativistic dynamics, 
if the position variables that are used 
to describe a system of charged interacting particles 
are the coordinates associated to a Lorentz frame~
\footnote{For coordinates belonging to a Lorentz frame 
we mean coordinates which transform as linear representation of 
the Poincar\`e group~\cite{DS}.}, then 
all higher time derivatives must appear in the Lagrangian~\cite{K65}. 
To get an ``ordinary'' Lagrangian it is necessary to introduce 
canonical position variables different from the Lorentz 
coordinates~\cite{K65}.
At 2PC order the acceleration dependent Lagrangian was originally 
derived by Golubenkov and Smorodinskii~\cite{GS56}. 
If one eliminates in that Lagrangian the higher time 
derivatives by using the equation of motion of lower orders then,
as pointed out in~\cite{K65,D82}, one does not obtain the correct equations of motion 
in a Lorentz frame. To eliminate correctly the accelerations 
one can use the method of ``redefinition of position variables'', 
introduced by Damour and Sch\"afer in \cite{DS},  
which consists in appealing to a contact transformation 
induced by a change of coordinates from the Wheeler-Feynman 
coordinate system (Lorentz frame)~\cite{WF49} to a well 
defined asymptotically inertial frame \cite{S84}. 
More explicitly, the acceleration dependent Lagrangian at 
2PC order is given by~\cite{DS}:
\beq
\label{r2.1}
\tilde{\cL}(\bz_1,\bz_2,\bv_1,\bv_2,\ba_1,\ba_2) = \tilde{\cL}_0 + 
\frac{1}{c^2}\,\tilde{\cL}_2 + \frac{1}{c^4}\,\tilde{\cL}_4 \,,
\eeq
with
\bea
\label{r2.2}
\tilde{\cL}_0 &=& 
\frac{1}{2}m_1\,\bv_1^2 + \frac{1}{2}m_2\,\bv_2^2 - 
\frac{e_1\,e_2}{R}\,,\\
\label{r2.3}
\tilde{\cL}_1 &=& \frac{1}{8}m_1\,\bv_1^4 + \frac{1}{8}m_2\,\bv_2^4 + 
\frac{e_1\,e_2}{2 R}\,\left [ \bv_1 \cdot \bv_2 + 
(\tilde{\bn} \cdot \bv_1)(\tilde{\bn} \cdot \bv_2)\right ]\,,\\
\label{r2.4}
\tilde{\cL}_4 &=& \frac{1}{16}m_1\,\bv_1^6 + \frac{1}{16}m_2\,\bv_2^6 - 
\frac{e_1\,e_2}{8}\,\left \{ 
R\,\left [ 3(\ba_1 \cdot \ba_2) - (\tilde{\bn} \cdot \ba_1)\,(\tilde{\bn} \cdot \ba_2) \right ]
+ 2 \left [ (\bv_1 \cdot \ba_2)\,(\tilde{\bn} \cdot \bv_1) \right . \right . \nonumber \\
&& \left . \left . - (\bv_2 \cdot \ba_1)\,(\tilde{\bn} \cdot \bv_2) \right ] + 
(\tilde{\bn} \cdot \ba_1)\, \left [ \bv_2^2 - (\tilde{\bn} \cdot \bv_2)^2 \right ]
-  (\tilde{\bn} \cdot \ba_2)\, \left [ \bv_1^2 - (\tilde{\bn} \cdot \bv_1)^2 \right ]\right .
\nonumber \\
&& \left . + \frac{1}{R}\,\left [ \bv_1^2\,\bv_2^2 -2 (\bv_1 \cdot \bv_2)^2 - 
\bv_1^2\,(\tilde{\bn} \cdot \bv_2)^2 -  \bv_2^2\,(\tilde{\bn} \cdot \bv_1)^2 
+ 3 (\tilde{\bn} \cdot \bv_1)^2\,(\tilde{\bn} \cdot \bv_2)^2 \right ]\right \}\,, 
\eea
where $\bR = \bz_1 - \bz_2$, $\tilde{\bn} = \bR/R$, $\bv_i = \dot{\bz}_i$ and 
$\ba_i = \dot{\bv}_i$.
In \cite{DS}, Damour and Sch\"afer after having critically discussed and clarified 
the various results previously derived in the literature \cite{CED},  
worked out the contact transformations, 
\bea
\bq_1 = \bz_1 - \frac{1}{c^4}\,\frac{e_1\,e_2}{4\,m_1}\,\left \{
(\tilde{\bn}\cdot \bv_2)\,\bv_2 + \tilde{\bn} \,\left [
\frac{1}{2}\,( (\tilde{\bn}\cdot \bv_2)^2 - \bv_2^2 ) + 
\frac{e_1\,e_2}{m_2\,R} \right ] \right \}\,,\\
\bq_2 = \bz_2 + \frac{1}{c^4}\,\frac{e_1\,e_2}{4\,m_2}\,\left \{
(\tilde{\bn}\cdot \bv_1)\,\bv_1 + \tilde{\bn} \,\left [
\frac{1}{2}\,( (\tilde{\bn}\cdot \bv_1)^2 - \bv_1^2 ) + 
\frac{e_1\,e_2}{m_1\,R} \right ] \right \}\,,
\eea
which allow to eliminate the accelerations appearing 
in Eqs.~(\ref{r2.2})--(\ref{r2.4}).  Hence, the final acceleration 
independent Lagrangian at 2PC order is given by \cite{DS}:
\beq
\label{2.1}
\cL(\bq_1,\bq_2,\bqd_1,\bqd_2)
= \cL_0 + \frac{1}{c^2}\,\cL_2 + \frac{1}{c^4}\,\cL_4 \,,
\eeq
with
\bea
\label{2.2}
\cL_0 &=& \frac{1}{2}m_1\,\bqd_1^2 + \frac{1}{2}m_2\,\bqd_2^2 - 
\frac{e_1\,e_2}{q}\,,\\
\label{2.3}
\cL_2 &=& \frac{1}{8}m_1\,\bqd_1^4 + \frac{1}{8}m_2\,\bqd_2^4 + 
\frac{e_1\,e_2}{2 q}\,\left [ \bqd_1 \cdot \bqd_2 + 
(\bn \cdot \bqd_1)(\bn \cdot \bqd_2)\right ]\,,\\
\label{2.4}
\cL_4 &=& \frac{1}{16}m_1\,\bqd_1^6 + \frac{1}{16}m_2\,\bqd_2^6 - 
\frac{e_1\,e_2}{8 q}\,\left \{ \bqd_1^2\,\bqd_2^2 - 2 (\bqd_1 \cdot \bqd_2)^2 + 
3 (\bn \cdot \bqd_1)^2 (\bn \cdot \bqd_2)^2 \right . \nonumber \\
&& \left . - (\bn \cdot \bqd_1)^2\,\bqd_2^2 
 - (\bn \cdot \bqd_2)^2\,\bqd_1^2 
+ \frac{e_1\,e_2}{m_2\,q}\, \left [ \bqd_1^2 - 
3 (\bn \cdot \bqd_1)^2 \right ] \right . \nonumber \\ 
&& \left. + \frac{e_1\,e_2}{m_1\,q}\, \left [ \bqd_2^2 
- 3 (\bn \cdot \bqd_2)^2 \right ] 
- \frac{2(e_1 \,e_2)^2}{m_1\,m_2\,q^2} \right \} \,,
\eea
where $\bq = \bq_1 - \bq_2$ and $\bn = \bq/q$. 
Applying the Legendre transformation to $\cL$, we derive 
(in full agreement with \cite{DS})
\beq
\label{2.5}
\cH(\bq_1,\bq_2,\bp_1,\bp_2) = \cH_0 + \frac{1}{c^2}\,\cH_2 + \frac{1}{c^4}\,\cH_4 \,,
\eeq
where 
\bea
\label{2.6}
\cH_0&=& \frac{1}{2}\,\left ( \frac{\bp_1^2}{m_1} + 
\frac{\bp_2^2}{m_2} \right )+ \frac{e_1\,e_2}{q}\,, \\
\label{2.7}
\cH_2 &=& -\frac{1}{8}\,\left ( \frac{\bp_1^4}{m_1^3} + 
\frac{\bp_2^4}{m_2^3} \right ) - \frac{e_1\,e_2}{2 m_1\,m_2\,q}\,
\left [ \bp_1 \cdot \bp_2 + (\bn \cdot \bp_1) (\bn \cdot \bp_2) \right ]\,,\\
\label{2.8}
\cH_4 &=& \frac{1}{16}\left ( \frac{\bp_1^6}{m_1^5} + \frac{\bp_2^6}{m_2^5} \right )
+ \frac{e_1\,e_2}{m_1\,m_2\,q}\, \left \{ 
\frac{3(\bn \cdot \bp_1)^2\,(\bn \cdot \bp_2)^2}{8m_1\,m_2} 
- \frac{\bp_1^2\,(\bn \cdot \bp_2)^2}{8 m_1\,m_2} 
- \frac{\bp_2^2\,(\bn \cdot \bp_1)^2}{8 m_1\,m_2} \right . \nonumber \\
&& \left. + \frac{1}{4}\left [ (\bn \cdot \bp_1)\,(\bn \cdot \bp_2) + 
(\bp_1 \cdot \bp_2) \right ]\,\left (\frac{\bp_1^2}{m_1^2} + 
\frac{\bp_2^2}{m_2^2} \right ) -\frac{(\bp_1 \cdot \bp_2)^2}{4m_1\,m_2} + 
\frac{\bp_1^2\,\bp_2^2}{8 m_1\,m_2} \right . \nonumber \\
&& \left . + \frac{e_1\,e_2}{q}\,\left ( \frac{\bp_1^2}{m_1} + 
\frac{\bp_2^2}{m_2} \right ) - \frac{(e_1\,e_2)^2}{4q^2} \right \}\,. 
\eea
Let us denote 
\beq 
\label{2.9}
M = m_1 + m_2\,, \quad \quad \mu = \frac{m_1\,m_2}{M}\,, \quad \quad 
\nu = \frac{\mu}{M}\,,
\eeq
where the parameter $\nu$ takes values between 0 and 1/4, corresponding 
to the test mass limit and the equal mass case, respectively.
Henceforth, we shall limit to the dynamics of the bound states generated 
by the two charged bodies, therefore $e_1\,e_2 < 0$ and we pose 
the coupling constant $\alpha = -e_1\,e_2 >0$. In the center 
of mass frame we have $\bP = \bp_1 = - \bp_2$ and 
introducing the following reduced variables
\beq
\label{2.10}
\widehat{\cal H} = \frac{\cH}{\mu}\,, \quad \quad \bp = \frac{\bP}{\mu}\,,
\quad \quad \widehat{t} = \frac{\mu\,t}{\alpha} \,, 
\quad \quad r = \frac{\mu\,q}{\alpha}\,,
\eeq
we can re-write the Hamiltonian, Eq.~(\ref{2.5}), in the more convenient form
\bea
\label{2.11}
\widehat{\cH}(\br,\bp) &=&  \frac{1}{2}\,\bp^2 - \frac{1}{r} - 
\frac{1}{8c^2}\,(1 -3\nu)\,{\bp^4} - 
\frac{1}{2c^2}\,\frac{\nu}{r}\,
[\bp^2 + (\bn \cdot \bp)^2] \nonumber \\
&& - \frac{1}{8c^4}\frac{1}{r}\,
\left [ 3\nu^2\,(\bn \cdot \bp)^4 + \nu\,\left (3\nu - 2\right )\,\bp^4
+ 2\nu\,\left (\nu - 1\right )\,\bp^2\,(\bn \cdot \bp)^2 \right ] 
\nonumber \\
&& + \frac{1}{16 c^4}\,(1 - 5\,\nu + 5\,\nu^2)\,{\bp^6} +
\frac{1}{4 c^4}\frac{\nu}{r^2}\,{\bp^2} 
+ \frac{1}{4 c^4}\,\frac{\nu}{r^3}\,.
\eea
The above Hamiltonian is invariant under time translations and space rotations.  
We denote the two conserved quantities, that is the centre-of-mass 
non-relativistic energy and angular momentum, by
\beq
\label{2.12}
\widehat{\cH}(\br,\bp) = \widehat{{\cal E}}^{\rm NR} = 
\frac{{\cal E}_{\rm c.m.}^{\rm NR}}{\mu}\,, 
\quad \quad \br \wedge \bp = \bj = \frac{\bcJ_{\rm c.m.}}{\alpha}\,.
\eeq
In the following we pose ${\cal E}^{\rm NR} \equiv {\cal E}_{\rm c.m.}^{\rm NR}$ 
and $\bcJ \equiv \bcJ_{\rm c.m.}$. Using the Hamilton-Jacobi formalism, we can summarize in a 
coordinate-invariant manner the two-charge dynamics  
by evaluating the ``energy-levels'' of the system.  
Introducing the reduced Hamilton principal-function $\widehat{S}$, 
defined by $(\partial \widehat{S}/\partial \br) = \bp$, separating 
the time and angular coordinates and restricting to the planar motion, 
we can write 
\beq
\label{2.13}
\widehat{S} = -\widehat{\cE}^{\rm NR}\,\widehat{t} + j\,\vphi + \widehat{S}_r(r,
\widehat{\cE}^{\rm NR},j)\,.
\eeq
Solving the Hamilton-Jacobi equation $\widehat{\cH}(\br,\bp) = 
\widehat{{\cal E}}^{\rm NR}$ with respect 
to $(d \widehat{S}_r/d r) = p_r = \bn \cdot \bp$, 
using $\bp^2 = (\bn \cdot \bp)^2 + \bj^2/r^2$, we get  
\beq 
\label{2.14}
\widehat{S}_r(r,\widehat{\cE}^{\rm NR},j) = \int dr\,\sqrt{{\cal R}(r,\widehat{\cE}^{\rm NR},j)}\,,
\eeq
where ${\cal R}$ is a polynomial of the fifth order in $1/r$, explicitly given by:
\beq
\label{2.15}
{\cal R}(r,\widehat{\cE}^{\rm NR},j) = A + \frac{2 B}{r} + 
\frac{C}{r^2} + \frac{D_1}{r^3} + \frac{D_2}{r^4} + 
\frac{D_3}{r^5}\,,
\eeq
with 
\bea
\label{2.16}
A &=& 2\widehat{\cE}^{\rm NR} + 
\frac{1}{c^2}\,(1 -3\nu)\,(\widehat{\cE}^{\rm NR})^2 + \frac{1}{c^4}\,\nu\,(4\nu-1)\,
(\widehat{\cE}^{\rm NR})^3\,,\\
\label{2.17}
B &=& 1 + \frac{1}{c^2}\,(1-\nu)\,\widehat{\cE}^{\rm NR} + 
\frac{1}{c^4}\,\frac{\nu}{2}\,\left ( 2\nu -1\right )\,(\widehat{\cE}^{\rm NR})^2\,,\\
\label{2.18}
C &=& -j^2 + \frac{1}{c^2}\,(1 + \nu)\,,\\
\label{2.19}
D_1 &=& -\frac{1}{c^2}\,\nu\,j^2 - 
\frac{1}{c^4}\,\nu^2\,j^2\,\widehat{\cE}^{\rm NR} 
+ \frac{1}{c^4}\,\frac{\nu}{2}\,\left ( 4\nu -1 \right )\,,\\
\label{2.20}
D_2 &=& -\frac{3}{c^4}\,\nu^2\,j^2\,,\\
\label{2.21}
D_3 &=& +\frac{3}{4c^4}\,\nu^2\,j^4\,.
\eea
For our purposes we need to compute the reduced radial action variable
\beq
\label{2.22}
i_r^{\rm real}(\widehat{\cE}^{\rm NR},j) = \frac{2}{2\pi} 
\int_{r_{\rm min}}^{r_{\rm max}} dr\,\sqrt{{\cal R}(r,\widehat{\cE}^{\rm NR},j)}\,.
\eeq
To evaluate the above integral we use the formula (3.9) 
of Ref.~\cite{DS88}, derived by performing a complex contour integration. 
The result for the radial action variable $\cI^{\rm real}_R = \alpha\,i^{\rm real}_r$  
reads:
\bea
\label{2.23}
\cI_R^{\rm real}(\cE^{\rm NR},\cJ)  &=& \frac{\alpha\,\mu^{1/2}}{\sqrt{-2\cE^{\rm NR}}}\,
\left [ 1 - \frac{1}{4}(\nu -3)\,\frac{\cE^{\rm NR}}{\mu\,c^2} - 
\frac{1}{32}\,(5 - 6\,\nu -3\,\nu^2)\,
\left (\frac{\cE^{{\rm NR}}}{\mu \,c^2}\right )^2 \right ]\nonumber \\
&& - \cJ + \frac{\alpha^2}{c^2\,\cJ}\,\left ( \frac{1}{2} - 
\frac{\nu}{2}\frac{\cE^{\rm NR}}{\mu\,c^2} \right )
+ \frac{1}{8}(1 - 6\nu)\,\frac{\alpha^4}{c^4\,\cJ^3}\,.
\eea
Finally, to get the ``energy-levels'' we solve the above 
equation in terms of the relativistic energy 
$ \cE^{\rm R} = \cE^{\rm NR} + M\,c^2$. Introducing the Delaunay 
action variable ${\cal N} = \cI_R^{\rm real} + \cJ$, we get:
\bea
\label{2.24}
\cE^{\rm R}({\cal N},\cJ) &=&  M\,c^2 - \frac{1}{2}\,\frac{\alpha^2\,\mu}{{\cal N}^2} + 
\frac{1}{c^2}\,\alpha^4\,\mu\, \left [ - \frac{1}{2}\frac{1}{\cJ\,{\cal N}^3} 
+ \frac{1}{8}(3 -\nu)\,\frac{1}{{\cal N}^4} \right ] + 
\frac{1}{c^4}\,\alpha^6\,\mu\,\left [ - \frac{3}{8}\frac{1}{\cJ^2\,{\cal N}^4} 
 \right. \nonumber \\
&& + \left. \frac{1}{16}\,(-5 + 3\nu -\nu^2) 
\frac{1}{{\cal N}^6} + \frac{1}{4}\left (3 - 2\nu \right )\frac{1}{\cJ\,{\cal N}^5}
+ \frac{1}{8}\left ( 6\nu - 1\right )\,
\frac{1}{\cJ^3\,{\cal N}^3} \right ]\,. 
\eea
At 0PC order we recover the well known result of the degeneracy 
of the energy-levels in the Coulomb problem.  
Let us observe that at 1PC order, identifying ${\cal N}/\hbar$ 
with the principal quantum-number and $\cJ/\hbar$ with 
the total angular-momentum quantum-number,  
we obtain that Eq.~(\ref{2.24}) gives, e.g., the correct 
bound-state energies of the singlet states of   
the positronium  \cite{BIZ70,T70} ($e_1 = -e_2$ and $m_1 = m_2$) 
in the (classical) limit $\cJ/\hbar \gg 1$.
Moreover, within the approximation $\cJ/\hbar \gg 1$, 
our method captures all the centrifugal barrier 
shifts that have to be added by hand 
in \cite{BIZ70}. However, we cannot recover from Eq.~(\ref{2.24})
the correct quantum energy-levels at 2PC level, because  
at this order radiation reaction effects should have been taken into account. 
Indeed, in electrodynamics they enter at 1.5PC order, with a dipole-type 
interaction. Only if we limit to systems with $e_1/m_1 = e_2/m_2$,  
we can postpone radiation reaction effects at the quadrupole order, 
which means at 2.5 PC level. In the present work 
we are interested in the conservative part of the bound states dynamics, 
hence we do not make the restriction $e_1/m_1 = e_2/m_2$. 
The radiative corrections which contribute to the main part of 
the Lamb shift have been evaluated in \cite{T70}, using the quasi-potential 
approach, and are of the order $\alpha^5\,\log \alpha$. Corrections 
of the order $\alpha^5$, $\alpha^6$, $\alpha^6\,\log\alpha$ have also 
been partially obtained in the literature for some quantum 
bound states of positronium and muonium~\cite{Lit}.

\section{``Effective'' one-body description}
\label{sec3}
The basic idea of the present work is to map the ``real'' 
two-body dynamics, described in the previous section, to an 
``effective'' dynamics of a test particle of mass $m_0$ 
and charge $e_0$, moving in an external electromagnetic 
field. The action for the test particle is given by:
\beq
\label{3.1}
S_{\rm eff} = \int \left (-m_0\,c\,ds_0 + \frac{1}{c}\,e_0\,A^{\rm eff}_\mu(z)\,dz^\mu
\right )\,,
\eeq
where $A_{\rm eff}^\mu = (\Phi_{\rm eff},\bA_{\rm eff})$. It is straightforward  
to derive that the effective Hamiltonian satisfies the well known equation
\beq
\label{3.2}
\frac{(\cH_{\rm eff} - e_0\,\Phi_{\rm eff})^2}{c^2}=  
m_0^2\,c^2 + \left ( \bp - \frac{e_0}{c}\,\bA_{\rm eff} \right )^2\,.
\eeq
The effective electromagnetic field $A_{\rm eff}^\mu$ 
will be constructed in the form of an expansion in  the 
dimensionless parameter $\alpha_0/(m_0\,c^2\,R)$, 
where $\alpha_0 = e_0^2$ is the coupling constant and 
$\alpha_0/(m_0\,c^2)$ is the classical charge radius of $m_0$. Hence, we pose:
\bea
\label{3.3}
&& \Phi_{\rm eff}(R) = \frac{e_0\,\phi_0}{R}\,\left [ 1 + \phi_1\,\frac{\alpha_0}{m_0\,c^2\,R} + 
\phi_2\,\left (\frac{\alpha_0}{m_0\,c^2\,R}\right )^2 + \cdots \right ]\,,\\
&& 
\label{3.4}
\bA_{\rm eff}(R) = \frac{e_0\,\ba}{c\,R} \left [a_0 + a_1\,
\frac{\alpha_0}{m_0\,c^2\,R} + \cdots \right]\,,
\eea
where $\phi_0,\phi_1, \phi_2$ and $a_0,a_1$ are dimensionless 
parameters and $\ba$ is a vector with the dimension of a velocity.
All these unknown coefficients will be fixed by the matching between the 
``real'' and the ``effective'' description.
Note that, in the above equations the variable $R$ stands for the effective 
radial coordinate and differs from the real separation $R$ used 
in Sec.~\ref{sec2}. Moreover, in Eqs.~(\ref{3.3}), (\ref{3.4}) 
we have indicated only the terms we shall need up to 2PC order.

The dynamics of the one-body problem can be described, 
in a coordinate-invariant manner,  
in the Hamilton-Jacobi framework, by considering the ``energy-levels'' 
of the bound states of the particle $m_0$ in the external 
electromagnetic field. The Hamilton-Jacobi equation can be obtained 
from Eq.~(\ref{3.2}) posing $\cH_{\rm eff} = {\cal E}_0$ 
and introducing the Hamilton principal-function 
$\partial S_{\rm eff}/\partial \bR = \bp$. 
Limiting to the motion in the equatorial plane ($\theta = \pi/2$) we 
can separate the variables, writing 
\beq
\label{3.5}
S_{\rm eff} = - {\cal E}_0\,t + \cJ_0\,\varphi + S_R^{0}(R,\cE_0,\cJ_0)\,, 
\eeq
where ${\cal E}_0$ and $\cJ_0 \equiv |\bcJ_0|$ are the conserved energy and angular momentum 
defined by Eq.~(\ref{3.1}). The effective radial action variable 
reads
\beq
\label{3.6}
\cI_R^{\rm eff} = \frac{2}{2\pi} 
\int_{R_{\rm min}}^{R_{\rm max}} dR\,\frac{d S_R^0}{d R}\,.
\eeq
Like in the two-body description we can derive  
the ``energy-levels'' of the ``effective'' one-body 
problem. They can be written as~\footnote{Note that, if a vector potential is present,  
the energy-levels could also depend on the magnetic 
number $\cJ_z^0$. In the present paper when dealing 
with a vector potential (see Sec.~\ref{subsec3.2}) we shall assume that the source of 
the magnetic field is the angular momentum, hence 
the magnetic field will be perpendicular to the plane of motion. 
This choice implicitly assumes $\cJ_0 \equiv \cJ_z^0$.}:
\bea
\label{3.7}
{\cal E}_0 ({\cal N}_0\ , \cJ_0) &= & m_0 \, c^2 - \frac{1}{2} \, \frac{m_0 
\, \alpha_0^2}{{\cal N}_0^2} + 
\frac{1}{c^2}\,\alpha_0^4\,m_0\,
\left (\frac{{\cal E}_{3,1}}{\cJ_0\,{\cal N}^3_0} + \frac{{\cal E}_{4,0}}{{\cal N}_0^4} \right) 
\nonumber \\
&+ & \frac{1}{c^4}\,\alpha_0^6\,m_0\, \left [ \frac{{\cal E}_{3,3}}{\cJ_0^3\,{\cal N}^3_0} + 
\frac{{\cal E}_{4,2}}{\cJ_0^2\,{\cal N}_0^4} + \frac{{\cal E}_{5,1}}{\cJ_0\,{\cal N}_0^5} 
+ \frac{{\cal E}_{6,0}}{{\cal N}_0^6} \right ]\,, 
\eea
where ${\cal N}_0 = \cI_R^{\rm eff} + \cJ_0$ and ${\cal E}_{i,j}$ are combinations 
of the coefficients $\phi_0,\phi_1, \phi_2$ and $a_0,a_1$ 
given in Eqs.~(\ref{3.3}), (\ref{3.4}).

Let us now define the rules to match the ``real'' to the ``effective'' 
problem. Like in \cite{BD99}, we find very natural sticking with 
the following relations between the adiabatic invariants:
\beq
\label{3.8}
{\cal N} = {\cal N}_0 \,, \quad \quad \cJ = \cJ_0\,.
\eeq
However, the way the ``energy-levels'', Eq.~(\ref{2.24}) and Eq.~(\ref{3.7}), are 
related is more subtle. If we 
simply identify $\cE_0(\cJ_0,\cN_0) = \cE^{\rm R}(\cJ,\cN) + (m_0 -M)\,c^2$, 
and impose that the mass of the effective test particle 
coincides with the reduced mass, i.e. $m_0 = \mu$, 
we obtain that already at 1PC order it is 
impossible to reduce the two-body dynamics to a one-body description. 
Hence, following \cite{BD99} we assume that there is a one-to-one 
mapping between the ``real'' and the ``effective'' 
energy-levels of the general form:   
\beq
\label{3.9}
\frac{\cE_0^{\rm NR}}{m_0\,c^2} = \frac{\cE^{\rm NR}}{\mu\,c^2}\,
\left [ 1 + \alpha_1\,\frac{\cE^{\rm NR}}{\mu\,c^2} + 
\alpha_2\,\left (\frac{\cE^{\rm NR}}{\mu\,c^2} \right )^2
\right ]\,,
\eeq
where $\alpha_1$ and $\alpha_2$ are unknown coefficients that will 
be fixed by the matching. Given the aforesaid ``rules'',  
we shall investigate in the subsequent sections 
three feasible ways the mapping can be implemented. The diverse  
descriptions differ by the choice of the effective electromagnetic field and  
the spacetime metric.

\subsection{Effective scalar potential depending on the energy}
\label{subsec3.1}
In this section we study the possibility of reducing the 
two-body dynamics to a one-body one introducing, in the ``effective'' 
description, the scalar potential $\Phi_{\rm eff}$ 
displayed in Eq.~(\ref{3.3}),  
and assuming that the vector potential $\bA_{\rm eff}$ is zero. 
In this case the derivative of the radial Hamilton principal-function is given by:
\beq
\label{3.1.1}
\frac{d S_R^0}{d R} = 2m_0\,\cE_0^{\rm NR} -2m_0\,e_0\,\Phi_{\rm eff} - 
\frac{\cJ_0^2}{R^2} + \frac{(\cE_0^{\rm NR})^2}{c^2} + \frac{e_0^2\,\Phi_{\rm eff}^2}{c^2} 
-\frac{2e_0\,\cE_0^{\rm NR}\,\Phi_{\rm eff}}{c^2}\,,
\eeq
where we have introduced the non-relativistic energy $\cE^{\rm NR}_0 = 
\cE_0^{\rm R} - m_0\,c^2$. Plugging the above expression in Eq.~(\ref{3.6})  
we get:
\bea
\label{3.1.2}
&& \cI^{\rm eff}_R(\cE_0^{\rm NR},\cJ_0) = \frac{\alpha_0\,m_0^{1/2}}{\sqrt{-2\cE_0^{\rm NR}}}\, 
\left [ -\phi_0 - \frac{3\phi_0}{4}\,\frac{\cE_0^{\rm NR}}{m_0\,c^2}
+ \frac{5\phi_0}{32}\left (\frac{\cE_0^{\rm NR}}{m_0\,c^2} \right )^2 \right ] 
-\cJ_0 \nonumber \\
&& + \frac{\alpha_0^2}{\cJ_0\,c^2}\,\left [ 
\frac{\phi_0^2}{2} - \phi_0\,\phi_1 - \phi_0\,\phi_1\,
\frac{\cE_0^{\rm NR}}{m_0\,c^2} \right ] + 
\frac{1}{8}\frac{\alpha_0^4}{\cJ_0^3\,c^4}\,\left [ \phi_0^4 - 12\phi_0^3\,\phi_1
+ 8\phi_0^2\,\phi_2 + 4\phi_0^2\,\phi_1^2 \right]\,.
\eea
Identifying Eq.~(\ref{3.1.2}) with Eq.~(\ref{2.23}), 
assuming $m_0 = \mu$ and using Eqs.~(\ref{3.8}), (\ref{3.9}) we  
obtain the equations for the unknowns 
$\phi_0,\phi_1, \phi_2$, $a_0,a_1$ and $\alpha_1$ and $\alpha_2$. In particular, 
at 0PC order we have 
\beq
\label{3.1.3}
- \phi_0\,\alpha_0 = \alpha\,, 
\eeq
and we find quite natural to pose $\phi_0 = -1$, that is $e_0^2 = \alpha_0 = \alpha = -e_1\,e_2$.  
The equations at 1PC level are:  
\beq
\label{3.1.4}
-\phi_0\,\alpha_0\,(2\alpha_1 -3)= 
\alpha\,(\nu -3)\,, \quad \quad \alpha_0^2\,(\phi_0^2 -2\phi_0\,\phi_1) = \alpha^2\,,
\eeq
while at 2PC order they read: 
\bea
\label{3.1.5}
&& -\phi_0\,\alpha_0\,(5 -12\alpha_1 - 12\alpha_1^2 
+ 16\alpha_2) = \alpha\,(5 - 6\nu - 3\nu^2)\,,\\
\label{3.1.6}
&& \alpha_0^4(\phi_0^4 + 4\phi_0^2\,\phi_1^2 - 12\phi_0^3\,\phi_1 + 
8\phi_0^2\,\phi_2) = \alpha^4(1 - 6\nu)\,, \\
\label{3.1.7}
&& \phi_0\,\phi_1\,\alpha_0^2 = \frac{\nu}{2}\,{\alpha^2}\,. 
\eea
Let us notice that at 1PC order, Eq.~(\ref{3.1.4})
gives $\alpha_1 = \nu/2$ and $\phi_1 = 0$. 
Then at 2PC order one can solve Eqs.~(\ref{3.1.5}) and (\ref{3.1.6})
in terms of $\alpha_2$ and $\phi_2$, obtaining $\alpha_2 =0$ and $\phi_2 = -3\nu/4$,
but Eq.~(\ref{3.1.7}) is inconsistent.  
To solve this incompatibility we are obliged to introduce another parameter 
in the ``effective'' description.
A simple possibility is to suppose that the diverse coefficients that appear in the 
effective scalar potential depend on an external parameter $E_{\rm ext}$, having 
the dimension of an energy, that is: 
\bea
\label{3.1.8}
\phi_0(E_{\rm ext}) &=& \phi_0^{(0)} + \phi_0^{(2)}\,\frac{E_{\rm ext}}{m_0\,c^2} + 
\phi_0^{(4)}\,\left (\frac{E_{\rm ext}}{m_0\,c^2}\right )^2\,,\\
\label{3.1.9}
\phi_1(E_{\rm ext}) &=& \phi_1^{(0)} + \phi_1^{(2)}\,
\frac{E_{\rm ext}}{m_0\,c^2}\,, \\ 
\label{3.1.10}
\phi_2(E_{\rm ext}) &=& \phi_2^{(0)}\,.
\eea
We find that in order to implement the matching with the ``real'' 
description the parameter $E_{\rm ext}$ should be fixed equal to the ``effective'' 
non-relativistic energy, i.e.  $E_{\rm ext} \equiv \cE_0^{\rm NR}$. 
In more details, the introduction of an energy dependence in the coefficients 
$\phi_0, \phi_1, \phi_2$ reshuffles the $c^{-2}$ expansion of Eq.~(\ref{3.1.2}), 
modifying the Eqs.~(\ref{3.1.4})--(\ref{3.1.7}) and allowing to solve in many 
ways the constraint equations. The simplest solution 
is envisaged by requiring that the energy-dependence 
enters only at 2PC order in the coefficient $\phi_1$. 
In this case, the solution reads: 
\bea
\label{3.1.11}
&& \phi_0^{(0)} = -1\,,\quad \quad \phi_0^{(2)} = 0\,, \quad \quad \phi_0^{(4)} = 0\,, \\
\label{3.1.12}
&& \phi_1^{(0)} = 0\,,\quad \quad  \phi_1^{(2)} = -\frac{\nu}{2}\,,
\quad \quad \phi_2^{(0)} = - \frac{3}{4}\nu\,, \\
\label{3.1.13}
&& \alpha_1 = \frac{\nu}{2}\,,\quad \quad \alpha_2 =0\,.
\eea
To summarize, we have succeeded in mapping the two-body dynamics 
onto the one of a test particle of mass $m_0 = \mu$ moving in the 
external scalar potential:
\beq
\label{3.1.14}
\Phi_{\rm eff}(R,E_{\rm ext}) = -\frac{e_0}{R}\,
\left [ 1  - \frac{\nu}{2}\,\left ( \frac{E_{\rm ext}}{m_0\,c^2}\right )\,
\left ( \frac{\alpha_0}{m_0\,c^2\,R}\right ) 
-\frac{3\nu}{4}\,\left (\frac{\alpha_0}{m_0\,c^2\,R}\right )^2 \right ] \,,
\eeq
where $E_{\rm ext} \equiv \cE^{\rm NR}_0$.
We have found that the matching is implemented 
relating the ``real'' and ``effective'' energy-levels by the formula:
\beq
\label{3.1.15}
\frac{\cE_0^{\rm NR}}{m_0\,c^2} = \frac{\cE^{\rm NR}}{\mu\,c^2}\,
\left [ 1 + \frac{\nu}{2}\,\frac{\cE^{\rm NR}}{\mu\,c^2}\right ]\,,
\eeq
which, as noticed in \cite{BD99}, gives the following relation between the 
real total relativistic energy $\cE$ and the effective relativistic energy 
$\cE_0$:
\beq
\label{3.1.16}
\frac{\cE_0}{m_0\,c^2} \equiv \frac{\cE^2 - m_1^2\,c^4 - m_2^2\,c^4}{2m_1\,m_2\,c^4}\,.
\eeq
The above equation has a rather interesting 
property. In the limit $m_1 \ll  m_2$
the effective energy of the effective 
particle equals the energy of the particle 
1 in the rest frame of particle 2 (and reciprocally if 
$m_2 \ll m_1$).
Moreover, the result (\ref{3.1.16}) coincides 
with the one derived in Ref.~\cite{BIZ70} 
in the context of quantum electrodynamics. 
We find quite remarkable that our way of relating the ``real'' and 
``effective'' energy-levels agrees with 
the one introduced in \cite{BIZ70}.
Nevertheless, we consider the dependence on the energy of the 
effective scalar potential, Eq.~(\ref{3.1.14}), quite unsatisfactory, though 
envisaged by Todorov et al.~\cite{T70} in the quasi-potential approach. 
Indeed, in our context the presence of an external parameter in 
the scalar potential obscures  the nature of the mapping and complicates 
the possibility of incorporating radiation reaction effects.  
Certainly, this cannot be achieved straightforwardly  in the way 
suggested in \cite{BD00} for the gravitational case.

As a final remark, let us note that if we were using the effective 
description introduced in the quasi-potential approach 
by Todorov et al.~\cite{T70}, 
we should have considered a test particle with effective mass, $m_{\rm eff}$, and 
effective energy, ${\cal E}_{\rm eff}$, given by:
\beq
m_{\rm eff}({\cal E}_{\rm real}) = \frac{m_1\,m_2\,c^2}{{\cal E}_{\rm real}}\,,\quad \quad 
\cE_{\rm eff}\equiv \frac{\cE^2_{\rm real} - m_1^2\,c^4 - m_2^2\,c^4}{2{\cal E}_{\rm real}}\,.
\eeq
We have investigated the possibility of introducing 
an energy dependence in the effective mass of the test particle, 
but we found that, in this case, it is not possible 
to overcome the inconsistency in the matching equations 
that raised at 2PC order. 
A way out could be to introduce also an energy dependence in 
the effective coupling $\alpha_{\rm eff}$, but we find this possibility not very 
appealing.  

\subsection{Effective vector potential depending on the angular momentum}
\label{subsec3.2}
We have seen in the previous section that, at 2PC level, 
in order to cope with an inconsistency of the constraint equations,  
we were obliged  to introduce an external parameter in 
the coefficients of the scalar potential. 
In this section we shall investigate the possibility of overcoming   
the above inconsistency by introducing, in the ``effective'' description,  
a scalar potential $\Phi_{\rm eff}$, independent of any external 
parameter, and a vector potential $\bA_{\rm eff}$ which 
will depend on an external vector $\bJ_{\rm ext}$. 
In order to implement the matching, we have found that it is sufficient   
to limit to the following form of the vector potential 
(see Eq.~(\ref{3.4})):
\beq
\label{3.2.1}
\bA_{\rm eff} = \frac{e_0\,(\bJ_{\rm ext} \wedge \bR)}{m_0\,c\,R^3}\, \left [a_0 + a_1\,
\frac{\alpha_0}{m_0\,c^2\,R} + \cdots \right]\,,
\eeq
where $\bJ_{\rm ext}$ is supposed to be perpendicular to the plane 
of motion~\footnote{Note that, with this choice of the vector potential 
the magnetic field will be perpendicular to the plane of motion.}.
In the Hamilton-Jacobi framework, restricting to $\theta = \pi/2$, 
we have
\beq
\bp = \frac{\partial S_{\rm eff}}{\partial \bR} = \widehat{\be}_R\,\frac{\partial S_{\rm eff}}{\partial R} + 
\widehat{\be}_\varphi\,\frac{1}{R}\,\frac{\partial S_{\rm eff}}{\partial \varphi}\,,
\eeq
where $\widehat{\be}_R$ and $\widehat{\be}_\varphi$ are vectors of the orthonormal basis. 
Due to the particular choice of the vector $\bJ_{\rm ext}$ we made, 
the following equation holds: 
\beq
\bA_{\rm eff} = \frac{e_0\,J_{\rm ext}\,\widehat{\be}_{\varphi}}{m_0\,c\,R^2}\, \left [a_0 + a_1\,
\frac{\alpha_0}{m_0\,c^2\,R} + \cdots \right]\,,
\eeq
where $J_{\rm ext} = |\bJ_{\rm ext}|$. Finally, using $\partial S_{\rm eff}/\partial \varphi = \cJ_0$ 
(see Eq.~(\ref{3.5})), we get:
\beq
\bp \cdot \bA_{\rm eff} = \frac{e_0\,J_{\rm ext}\, \cJ_0}{m_0\,c\,R^3}\,\left [a_0 + 
a_1\,\frac{\alpha_0}{m_0\,c^2\,R} + \cdots \right ] \,,
\quad \quad \bA_{\rm eff}^2 = \frac{e_0^2\,J_{\rm ext}^2\,a_0^2}{m_0^2\,c^2\,R^4} + \cdots \,.
\eeq
Note the crucial fact that, with the very special choice of the 
vector potential we made, $\bp \cdot \bA_{\rm eff}$ does not depend on $\bp_R$.
Plugging the above expressions in the Hamilton-Jacobi equation, 
Eq.~(\ref{3.2}), with $\cH_{\rm eff} = \cE^{\rm NR}_0 + m_0^2\,c^2$ we obtain:
\bea
\label{3.2.3}
\frac{d S_R^0}{d R} &=& 2m_0\,\cE_0^{\rm NR} -2m_0\,e_0\,\Phi_{\rm eff} - 
\frac{\cJ_0^2}{R^2} + \frac{(\cE_0^{\rm NR})^2}{c^2} + \frac{e_0^2\,\Phi^2_{\rm eff}}{c^2}
-\frac{2e_0\,\cE_0^{\rm NR}\,\Phi_{\rm eff}}{c^2} \nonumber \\
&& + \frac{2\cJ_0\,J_{\rm ext}}{R^2}\,\left [a_0\,\frac{\alpha_0}{m_0\,c^2\,R} + 
a_1\,\left (\frac{\alpha_0}{m_0\,c^2\,R} \right )^2\right ] 
- \frac{J_{\rm ext}^2}{R^2}\,a_0^2\,\left (\frac{\alpha_0}{m_0\,c^2\,R} \right )^2\,,
\eea
where $\Phi_{\rm eff}$ is given by Eq.~(\ref{3.3}).
Evaluating the radial action variable (see Eq.~(\ref{3.6})) we finally get:
\bea
\label{3.2.4}
&& \cI^{\rm eff}_R(\cE_0^{\rm NR},\cJ_0,J_{\rm ext}) 
= \frac{\alpha_0\,m_0^{1/2}}{\sqrt{-2\cE_0^{\rm NR}}}\, 
\left [ -\phi_0 - \frac{3\phi_0}{4}\,\frac{\cE_0^{\rm NR}}{m_0\,c^2}
+ \frac{5\phi_0}{32}\left (\frac{\cE_0^{\rm NR}}{m_0\,c^2} \right )^2 \right ] 
-\cJ_0 + \frac{\alpha_0^2}{\cJ_0\,c^2}\,\left [ 
\frac{\phi_0^2}{2} \right . \nonumber \\
&& \left . - \phi_0\,\phi_1 -\phi_0\,a_0\,\frac{J_{\rm ext}}{\cJ_0}
+ \frac{\cE_0^{\rm NR}}{m_0\,c^2}\,\left (- \phi_0\,\phi_1 - \phi_0\,a_0\,
\frac{J_{\rm ext}}{\cJ_0} + a_1\,\frac{J_{\rm ext}}{\cJ_0} 
+ a_0^2\, \frac{J^2_{\rm ext}}{\cJ^2_0} \right ) \right ] + 
\frac{1}{8}\frac{\alpha_0^4}{\cJ_0^3\,c^4}\,\left [ \phi_0^4 
\right . \nonumber \\
&& \left . - 12\phi_0^3\,\phi_1 
+ 8\phi_0^2\,\phi_2 + 4\phi_0^2\,\phi_1^2 + 24\phi_0^2\,\phi_1\,a_0\,\frac{J_{\rm ext}}{\cJ_0}
-12\phi_0^3\,a_0\,\frac{J_{\rm ext}}{\cJ_0} + 12\phi_0^2\,a_1\,\frac{J_{\rm ext}}{\cJ_0}
 + 24\phi_0^2\,a_0^2\,\frac{J^2_{\rm ext}}{\cJ^2_0}\right ]\,. \nonumber \\
\eea
Let us impose that the above equation coincides 
with the analogous expression for the ``real'' description, given by Eq.~(\ref{2.23}).  
Assuming $m_0 = \mu$ and using Eqs.~(\ref{3.8}), (\ref{3.9}) we  
derive the new constraint equations to be satisfied. 
At 0PC order we still have $-\phi_0\,\alpha_0 = \alpha$, and we pose $\phi_0 = -1$,
while at 1PC level we get:
\beq
\label{3.2.5}
-\phi_0\,\alpha_0\,(2\alpha_1 -3)= 
\alpha\,(\nu -3)\,, \quad \quad \alpha_0^2\,\left (\phi_0^2 - 
2\phi_0\,\phi_1 -2\phi_0\,a_0 \, \frac{J_{\rm ext}}{\cJ_0} \right ) = \alpha^2\,.
\eeq
The first equation in (\ref{3.2.5}) gives $\alpha_1 = \nu/2$, while 
the second one is automatically satisfied if we 
make the rather natural requirement
that either the Coulomb potential does not have any correction 
at 1PC order ($\phi_1 = 0$) or the vector potential enters 
only at the next Coulombian order ($a_0 = 0$).
Finally, the 2PC order constraints read: 
\bea
\label{3.2.6}
&&- \phi_0\,\alpha_0\,(5 -12\alpha_1  -12\alpha_1^2 
+ 16\alpha_2) = \alpha\,(5 - 6\nu - 3\nu^2)\,,\\
\label{3.2.7}
&&\alpha_0^4\,\left (\phi_0^4 + 4\phi_0^2\,\phi_1^2 - 12\phi_0^3\,\phi_1 + 
8\phi_0^2\,\phi_2 +24\phi_0^2\,\phi_1\,a_0\,\frac{J_{\rm ext}}{\cJ_0}\right .
 \nonumber \\
&&\left . -12 \phi_0^3\,a_0\,\frac{J_{\rm ext}}{\cJ_0} + 24\phi_0^2\,a_0^2\,
\frac{J^2_{\rm ext}}{\cJ_0^2}+ 12\phi_0^2\,a_1\,\frac{J_{\rm ext}}{\cJ_0} \right )
 = \alpha^4(1 - 6\nu)\,, \\
&& \alpha_0^2\,\left (\phi_0\,\phi_1 + \phi_0\,a_0\,\frac{J_{\rm ext}}{\cJ_0}
 - a_1\,\frac{J_{\rm ext}}{\cJ_0} - a_0^2\,\frac{J^2_{\rm ext}}{\cJ_0^2} \right )  = \nu\,\frac{\alpha^2}{2}\,.
\label{3.2.8}
\eea
Plugging the results obtained at 0PC 
and 1PC order in Eqs.~(\ref{3.2.6})--(\ref{3.2.8})
and assuming that the external vector $J_{\rm ext}$ 
coincides with the constant of motion $\cJ_0$, 
we end up with the unique, rather simple solution:
\beq
\label{3.2.9}
\phi_2 =0\,, \quad \quad a_1 = - \frac{\nu}{2}\,,\quad \quad \alpha_2 = 0\,.
\eeq
In conclusion, in this Section we have obtained that at 2PC order 
it is possible to reduce the two-charge dynamics to the one of a test particle moving 
in an effective electromagnetic field described by a Coulomb potential
$\Phi_{\rm eff}(R) = -{e_0}/{R}$
and a vector potential dependent on the external vector $\bJ_{\rm ext}$ ($ \equiv \bcJ_0$):
\beq
\label{3.2.10}
\bA_{\rm eff}(R,J_{\rm ext}) = -\frac{\nu}{2}\,\frac{e_0\,\alpha_0}{m_0^2\,c^3}\,
\frac{(\bJ_{\rm ext} \wedge \bR)}{R^4}\,.
\eeq
Moreover, quite remarkably, we have found, under rather natural 
assumptions, that the one-to-one mapping between 
the ``real'' and the ``effective'' energy-levels is still 
given by the formula (\ref{3.1.16}). 
However, as already discussed at the end of the previous section, the fact 
that the electromagnetic field still has to depend on external parameters 
is not very desirable. In the next section we shall investigate 
a feasible way out.

\subsection{Effective metric}
\label{subsec3.3}
So far we have seen that in order to succeed in reducing the two-body
dynamics onto a one-body description we were obliged to introduce 
external parameters, which have been identified either with the 
energy or the angular momentum of the test particle $m_0$. This result is not 
very appealing, especially when we want to incorporate 
radiation reaction effects. 
A possible way out would be to relax the hypothesis that in the 
one-body description the test particle move in a flat spacetime.
The effective spacetime metric should be viewed as an effective way of describing 
the global exchange of energy between the two charged particles 
in the ``real'' description.

The most general spherical symmetric metric written in Schwarzschild gauge has 
the form: 
\beq
\label{3.3.1}
d s_{\rm eff}^2 = -A(R)\,c^2\,dt^2 + B(R)\,dR^2 + R^2\,(d\theta^2 + 
\sin \theta^2\,d \varphi^2)\,,
\eeq
where the coefficients $A(R)$ and $B(R)$ are given as an expansion 
in the dimensionless parameter $\alpha_0/(m_0\,c^2\,R)$, that is:
\bea
\label{3.3.2}
&& A(R) = 1 + A_1\,\frac{\alpha_0}{m_0\,c^2\,R} + 
A_2\,\left (\frac{\alpha_0}{m_0\,c^2\,R} \right )^2 + 
A_3\,\left (\frac{\alpha_0}{m_0\,c^2\,R} \right )^3 +   
\cdots\,, \\
\label{3.3.3}
&& B(R) = 1 + {B_1}\,\frac{\alpha_0}{m_0\,c^2\,R}  
+ B_2\,\left (\frac{\alpha_0}{m_0\,c^2\,R} \right )^2 + \cdots\,.
\eea
The reduction, from the two-body problem to the one-body one, can 
simply be implemented assuming that in the 
``effective'' description only the scalar potential $\Phi_{\rm eff}$ is different 
from zero. In this case the derivative of the Hamilton principal-function 
reads:
\beq
\label{3.3.4}
\frac{d S_R^0}{d R} = \frac{B(R)}{c^2\,A(R)}\,(\cE_0 + 
m_0\,c^2 - e_0\,\Phi_{\rm eff})^2 - \frac{B(R)}{R^2}\,\cJ_0^2 - B(R)\,m_0^2\,c^2\,,
\eeq
and for the radial action variable we derive:
\bea
\label{3.3.5}
\cI^{\rm eff}_R(\cE^{\rm NR}_0,\cJ_0) &=& 
\frac{\alpha_0\,m_0^{1/2}}{\sqrt{-2\cE_0^{\rm NR}}}\, 
\left [ {\cal A}  + {\cal B}\,\frac{\cE_0^{\rm NR}}{m_0\,c^2}
+ {\cal C}\,\left (\frac{\cE_0^{\rm NR}}{m_0\,c^2} \right )^2 \right ] 
-\cJ_0 \nonumber \\
&& + \frac{\alpha_0^2}{\cJ_0\,c^2}\,\left [ {\cal D} + {\cal E}\,
\frac{\cE_0^{\rm NR}}{m_0\,c^2} \right ] + \frac{\alpha_0^4}{\cJ_0^3\,c^4}\,{\cal F}\,,
\eea
where the various coefficients can be written explicitly as:
\bea
\label{3.3.6}
{\cal A} &=& - \phi_0 - \frac{1}{2}A_1\,, \\
\label{3.3.7}
{\cal B} &=& - \frac{3}{4}\phi_0 + \left ( B_1 - \frac{7}{8}A_1 \right )\,,\\
\label{3.3.8}
{\cal C} &=& \frac{5}{32} \phi_0 + \left ( \frac{B_1}{4} - \frac{19}{64}A_1 \right )\,,\\
\label{3.3.9}
{\cal D} &=& \phi_0\,\left ( -\phi_1 - \frac{B_1}{2} + A_1 \right ) + 
\frac{1}{2}\phi_0^2 - \frac{1}{4}A_1\,B_1 + 
\frac{A_1^2}{2} - \frac{A_2}{2}\,,\\
\label{3.3.10}
{\cal E} &=& \phi_0\,\left ( -\phi_1 + A_1 - \frac{B_1}{2} \right ) + 
A_1^2 - A_2 - \frac{1}{2}A_1\,B_1 - \frac{B_1^2}{8} + \frac{B_2}{2}\,,\\
\label{3.3.11}
{\cal F} &=& \frac{1}{64}\,\left (24A_1^4 - 48A_1^2\,A_2 + 8A_2^2 + 16A_1\,A_3 - 8A_1^3\,B_1 
+ 8 A_1\,A_2\,B_1 - A_1^2\,B_1^2 + 4A_1^2\,B_2 \right ) \nonumber \\
&& + \frac{\phi_0}{16}\,\left (-16\phi_1\,A_1^2 + 24A_1^3 + 8\phi_2\,A_1 + 8\phi_1\,A_2 
- 32A_1\,A_2 + 8A_3 + 4\phi_1\,A_1\,B_1  \right .\nonumber \\
&& \left . - 8A_1^2\,B_1 + 4A_2\,B_1 - A_1\,B_1^2 +4A_1\,B_2 \right ) 
+ \frac{\phi_0^3}{4}\left (- 6\phi_1 + 4A_1 - B_1 \right ) + \frac{\phi_0^4}{8} \nonumber \\
&& + \frac{\phi_0^2}{16} (8\phi_1^2 -40\phi_1\,A_1 + 32A_1^2 + 16\phi_2 
-20 A_2 + 8\phi_1\,B_1 -10 A_1\,B_1 - B_1^2 + 4B_2)\,.
\eea
The above expressions coincide with the ones 
obtained in pure general relativity \cite{BD99}, 
once the limit $\phi_0 \rightarrow 0$ is considered and 
$\alpha_0$ is identified with the analogous quantity 
in the gravitational case, i.e. with $G m_1 m_2$ ($G$ 
is the Newton constant).
Let us now equate the ``real'', Eq.~(\ref{2.23}) 
and the ``effective'', Eq.~(\ref{3.3.5}),
radial action variables, assuming that the following relations 
hold: $\cJ_0 = \cJ$, $m_0 = \mu$ and Eq.~(\ref{3.9}).  
At 0PC order we get the constraint  $\alpha_0\,(-\phi_0 - A_1/2) = \alpha$ 
which can be naturally fulfilled imposing that $A_1 =0$ and posing $\phi_0 = -1$, 
as above. At 1PC level we derive:
\bea
\label{3.3.12}
&& -2\phi_0\,\alpha_0\,(2\alpha_1-3) + \alpha_0\,(7A_1 -8B_1 -2A_1\,\alpha_1) = 
2\alpha\,(\nu-3)\,,\\
\label{3.3.13}
&& 2\alpha_0^2\,(\phi_0^2 + \phi_0\,(2A_1-B_1-2\phi_1)) + 
\alpha_0^2\,(2A_1^2 -2A_2 - A_1\,B_1) = 2\alpha^2\,.
\eea
If we demand that at this order the scalar potential and the effective metric 
do not differ from the Coulomb potential and the flat spacetime metric, respectively, 
i.e. we pose $ \phi_1 = 0, A_2 = 0, B_1 = 0$, 
we find that Eq.~(\ref{3.3.12}) gives $\alpha_1 = \nu/2$ 
while Eq.~(\ref{3.3.13}) is automatically satisfied. Inserting these values in the 
constraint equations at 2PC order and imposing that there are no corrections to the 
Coulomb potential at this order ($\phi_2 = 0$) we obtain the unique simple solution:
\beq
\label{3.3.14}
\alpha_2 = 0\,,
\quad \quad A_3 = \nu\,,\quad \quad B_2 = - \nu\,.
\eeq
Hence, we have found that with the introduction of an effective metric 
we are not obliged to introduce in the electromagnetic field 
any dependence on external 
parameters, neither the energy nor the angular momentum.
Moreover, up to 2PC order we find that there is no need 
of modifying the Coulomb scalar potential, 
i.e. $\Phi_{\rm eff}(R) = -e_0/R$ and the ``energy-levels''
of the real and ``effective'' description are still related by 
Eq.~(\ref{3.1.16}). Finally, the 
external spacetime metric is simply given by:
\beq
\label{3.3.15}
A(R) = 1 + \nu\,\left (\frac{\alpha_0}{m_0\,c^2\,R} \right )^3 \,,
\quad \quad B(R) = 1 -\nu\,\left (\frac{\alpha_0}{m_0\,c^2\,R} \right )^2\,.
\eeq

\section{Conclusions}
\label{sec4}
In this paper we have analysed the application 
of a new approach to studying the relativistic dynamics of 
the bound states of two classical charged particles, 
with comparable masses, interacting electromagnetically.
The key idea, originally introduced  
investigating the two-body problem in general relativity 
\cite{BD99}, has been to map the ``real'' two-body problem 
onto the one of a test particle 
moving in an external electromagnetic field.

We have found that the matching can be implemented 
imposing the following rather natural ``rules'': 
i) the adiabatic invariants 
${\cal N}$ and $\cJ$ in the two descriptions have 
to be identified; ii) the reduced mass of 
the ``real'' system, $\mu$, has to coincide with the mass 
of the effective particle, $m_0$, and  iii) 
the energy axis between the two problems has to be transformed. 
Let us note immediately that, a bottom-line 
of our results has been that, in all the three 
cases considered (see Sec.~\ref{subsec3.1}, 
\ref{subsec3.2} and \ref{subsec3.3}), 
we have found quite naturally that 
the energy axis, between the two descriptions, 
has to change in such a way that the effective energy of the effective 
particle coincides with the energy of the particle 
1 in the rest frame of particle 2 in the limit $m_1 \ll m_2$ 
(and vice versa) (see Eq.~(\ref{3.1.16})). 

Nevertheless, contrary to the results 
obtained in general-relativity \cite{BD99}, 
the requirements i), ii) and iii) envisaged above,  
do not fix uniquely the external electromagnetic field, with which 
the effective test particle $m_0$ interacts. 
In fact, we have found that, in order to overcome 
an inconsistency in the constraint equations which 
define the matching, we had to introduce an external 
parameter either in the scalar potential, Eq.~(\ref{3.1.14}),
or in the vector potential, Eq.~(\ref{3.2.10}). 
These parameters have to be identified 
with the non-relativistic energy and the angular momentum 
of the effective test particle $m_0$, respectively. As pointed 
out above and in Ref.~\cite{BD99}, 
the dependence of the effective electromagnetic field 
on some external parameter makes the  mapping between the two descriptions 
quite awkward and complicates the inclusion 
of radiation reaction effects. 
A possible solution of this issue is to relax
the hypothesis that the test particle moves 
in a flat spacetime. Indeed, in this case 
we have found that the conditions i), ii) and iii) fix 
rather naturally the external scalar potential and the
effective metric. They provide, up to 2PC order, 
an effective Coulomb potential and a rather simple $\nu$-deformed flat metric 
(see Eq.~(\ref{3.3.15})). 

Once the matching has been successfully defined, 
to have a complete knowledge of the ``real'' dynamics 
through the auxiliary ``effective'' one, 
we can construct, like in \cite{BD99}, the canonical transformation 
which relates the variables of the relative motion in the ``real'' description,  
to the coordinates and momenta of the test 
particle in the ``effective'' problem. However, 
this calculation goes beyond the scope of the present paper.

Finally, a last remark. In Sec.~\ref{subsec3.2}  
we have introduced a vector potential in the effective description 
in such a way that the source of the magnetic field 
is the angular momentum of the system. 
This study suggests the investigation,  
in the general relativity context~\cite{BD99}, 
of relaxing the hypothesis of 
mapping the ``real'' two-body dynamics onto 
the one of a test particle moving in a deformed 
Schwarzschild spacetime. Indeed, it could  
well be possible to match the two problems 
appealing to an effective deformed Kerr spacetime.

\section*{Acknowledgements}
It is a pleasure to thank Thibault Damour, Scott Hughes,  Gerhard Sch\"afer and Kip  Thorne 
for useful discussions and/or for comments on this manuscript.

This research is supported by the Richard C. Tolman Fellowship and 
by NSF Grant AST-9731698 and NASA Grant NAG5-6840.

\end{document}